\documentclass[10pt]{article}
\usepackage{graphicx}
\usepackage{verbatim}
\usepackage{amsmath}
\usepackage{amssymb}

\def\vp{\varphi}

\def\cO{{\mathcal O}}

\def\11{{\mathbb 1}}

\def\ri{{\rm i}}

\def\GeV{\,\rm{GeV}\,}
\def\GSM{\Gamma_{\rm SM}}

\def\beq{\begin{equation}}
\def\eeq{\end{equation}}
\def\bea{\begin{eqnarray}}
\def\eea{\end{eqnarray}}

\def\ri{\text{i}}

\begin{document}
\begin{center}
{\Large\bf A narrow scalar resonance  at 325 GeV?}

\vspace{0.5cm}
Krzysztof A. Meissner$^{1,2}$ and Hermann Nicolai$^2$

\vspace{0.5cm}
{\it $^1$ Faculty of Physics,
University of Warsaw, Ho\.za 69, Warsaw, Poland\\
$^2$ Max-Planck-Institut f\"ur Gravitationsphysik
(Albert-Einstein-Institut)\\
M\"uhlenberg 1, D-14476 Potsdam, Germany
}

\vspace{1cm}
\begin{minipage}[t]{12cm}{\small We propose to identify the excess of events with
four charged leptons at $E\sim 325\GeV$  seen by the CDF \cite{CDF}
and CMS \cite{CMS}  Collaborations with a new `sterile' scalar particle characterized by a very narrow resonance of the same height and branching ratios as the Standard Model Higgs boson, as predicted in the framework of the so-called Conformal
Standard Model \cite{MN}.
}
\end{minipage}
\end{center}

\vspace{0.5cm}

The long search for the Higgs particle culminated this summer with the announcement
by both LHC groups \cite{CMSATLASHiggs} of the discovery of a particle with mass of approximately $125\GeV$
that most probably corresponds to the Standard Model (=SM) Higgs boson. Prior to this
discovery, the search had produced several candidate events possibly hinting at new physics beyond the SM. In our opinion, the most striking of these was the excess
of $Z^0Z^0\to l^+ l^-  l^+ l^-$ events (four charged leptons) seen by the CDF Collaboration
\cite{CDF}, with an invariant mass of  almost precisely 325 GeV in all four cases, see Fig.1;
a smaller excess in this region was also reported by the CMS collaboration \cite{CMS}.
The excess of CDF events was treated as a statistical fluctuation \cite{CDF},
even though the probability of observing four such events
in a narrow energy band is estimated at $10^{-4}\,$--$\,10^{-5}$ in the SM.
This negative conclusion was mainly based on the fact that the two other decay channels that should accompany such events (two charged leptons plus missing energy, and two charged leptons plus two jets) did not show any excess at this energy in comparison with the expected rates for a SM Higgs boson of this mass.

In this letter we would like to point out a potential loophole in the argumentation
leading to the dismissal of these events. More specifically, we wish to call attention
to the possible existence of a new scalar particle in this mass range that would
show up as a {\em narrow} resonance (and in particular much narrower
than a SM Higgs resonance of that mass),  but {\em of the same height and
branching ratios} as the SM Higgs boson (i.e. about 70\% to $W^+W^-$, 30\% to
$Z^0 Z^0$ and below 0.1\% to $b\bar{b}$ and all remaining channels).
The existence of such a scalar particle with a mass close to the weak scale
is a prediction the so-called `Conformal
Standard Model' proposed in \cite{MN}. In addition to the usual Higgs
doublet this model necessarily contains an electroweak singlet (complex) scalar, which in the unbroken
phase couples only to right-chiral neutrinos and induces a Majorana mass term
for the right-chiral neutrinos via spontaneous symmetry breaking\footnote{Models with
  such an extra  scalar field were considered long ago (see \cite{BPY} and references
  therein), but there the relevant scale was always
  of the order of the assumed Majorana mass scale, i.e. above $10^{10}\GeV$.}.
Because of the mixing in the minimum of the potential the mass eigenstates of
the scalar particles not absorbed by the Brout-Englert-Higgs mechanism are mixtures
of the SM Higgs particle and the new scalar with some mixing angle $\theta$. As explained
below, this scenario would lead to two bumps in the cross section. Making the natural assumption that $\theta$ is small, the first bump would then coincide with the usual SM Higgs
maximum (now known to be at $M_1\sim125$ GeV)
with the height as computed for a SM Higgs of that mass, but with a slightly smaller
width $\Gamma_1=\GSM(M_1)\cos^2\theta$. The second bump, here
assumed to lie at about $M_2\sim 325$ GeV as suggested by the excess of CDF events
\cite{CDF}, would have {\em the same height} as the SM Higgs boson of that mass but its decay width would be reduced to $\Gamma_2=\GSM (M_2)\sin^2\theta$, and thus be very narrow in comparison with the width of a SM Higgs boson of that mass.  As can be seen from Fig.2
the SM Higgs decay width increases rapidly as a function of the Higgs mass, with
$\GSM (325{\rm\ GeV})\approx 20$ GeV. Taking $\theta=0.1$ as a plausible value, the
mixing would reduce the decay width down to 200 MeV.
This feature could explain the extremely small difference in the invariant masses
measured by CDF \cite{CDF}.

In the scenario proposed in \cite{MN}, the extra complex scalar field coupling
only to right-chiral neutrinos is required by the assumed conformal invariance of
the classical Lagrangian, which is a {\em non-supersymmetric} minimal extension
of the SM Lagrangian without explicit mass terms of any type, and in particular
no Majorana mass terms for the right-chiral neutrinos. The breaking of electroweak
symmetry and conformal symmetry (jointly leading to the generation of mass) is
assumed to occur via a Coleman-Weinberg type mechanism, such that the conformal
anomaly would be at the origin of mass generation. The very preliminary
analysis of \cite{MN} indicates that the mass of the second scalar is not
much above the mass of the usual SM Higgs boson and possibly below 500 GeV.
We note, however, that a reliable analysis of the symmetry breaking pattern
for a Coleman-Weinberg type potential with two independent scalar
fields would require the consistent incorporation of higher order corrections, whence
the mass values quoted in \cite{MN} should only be regarded as rough estimates.
In the absence of a detailed analysis we shall therefore simply {\em assume}
\begin{itemize}
\item Existence of a nontrivial stable minimum of the effective potential for the combined scalar
          sector exhibiting mixing of the SM Higgs with the new scalar; and
\item Viability of the model up to the Planck scale under the RG evolution of all
          couplings (as evaluated at the minimum of the effective potential), and compatibility
          of these values with the known Higgs mass.
\end{itemize}

The phase of the extra complex field $\phi(x)$ has been identified with the axion in \cite{MN2}, and its modulus $\vp(x) \equiv |\phi(x)|$  as the `heavy cousin' of the Higgs boson. The special feature of the model that is important here is the mixing between the Higgs field $H$ and the new scalar $\vp$. The former couples to SM particles in the standard way, whereas the new scalar is almost completely decoupled\footnote{Which is why we refer to this scalar particle
  as  `sterile'.}, since its coupling to the observable sector of the SM arises
only through the left-right neutrino mixing \cite{MN2}. More precisely, the mass eigenstates (with mass eigenvalues $M_1$ and $M_2$) are linear combinations of both fields:
\beq
\Phi_1 =  H\cos\theta+\vp\sin\theta \;\; , \quad
\Phi_2=-H\sin\theta+\vp\cos\theta
\eeq
With $\theta$ small, $\Phi_1$ would correspond to the recently discovered Higgs boson, while $\Phi_2$ is proposed here to cause the excess observed by CDF.
Then the amplitude of $Z^0Z^0$ production via the Higgs would be proportional
to ($\GSM$ is the Higgs decay rate in the SM\footnote{If the heavy neutrinos had mass $< \cO(M_2/2)$ the new scalar could also decay into a pair of heavy neutrinos. Then $\GSM$ and the branching ratio into missing energy should be appropriately enlarged.}
 \beq
{\cal A}\;\propto \;
\frac{\ri\cos^2\theta}{p^2-M_1^2+\ri M_1\GSM (M_1)\cos^2\theta}\;+\;
\frac{\ri\sin^2\theta}{p^2-M_2^2+\ri M_2\GSM (M_2)\sin^2\theta}
\eeq
Consequently, the decay widths are modified by factors $\cos^2\theta$ and $\sin^2\theta$, respectively, while the amplitude $|{\cal A}|$ is equal to the value of the amplitude
for a SM Higgs boson of the corresponding mass for $p^2 = M_1^2$ or $p^2 = M_2^2$,
because the dependence on the mixing angle cancels at the poles. When $\theta$ is small,
as we assume here, the second resonance is thus very narrow in comparison with the
expected width $\GSM (M_2)$. The event rates are directly obtained from
the known ones for the SM Higgs by multiplication with $\sin^2 \theta$.
Since the incoming partons in proton-proton
collisions range over a broad spectrum of energies and momenta, with low
initial probability of precisely `hitting' the narrow resonance, the emission of one
jet or two jets from the top quark triangle or the initial gluons producing the
resonance may be needed in order to `adjust' the energy to the required value
$s\approx M_2^2$. Therefore we expect that the products of the decay at
$s\approx M_2^2$ are generically accompanied by one or more jets, and have very
definite invariant mass in order to produce the enhancement in the
cross section (in contradistinction to a Higgs resonance of the same mass whose
invariant mass  distribution would be rather broad). The assumption of the existence of such a narrow resonance at 325 GeV makes the events seen by CDF much more
probable than if they were just due to statistical fluctuations of the background.
Nevertheless it is important to emphasize that the observation of four cases in one
bin at 325 GeV (out of total of eight events) still is a lucky  coincidence since this number exceeds the expected number of events even in the presence of a resonance at this energy, as one can easily see by comparing the estimated number of cases from the background with and without an assumed Higgs boson of mass 350 GeV \cite{CMS}. Therefore it would require a dedicated search with much more statistics than presently available at the LHC, especially if the resonance is very narrow, to prove (or disprove) the presence of such a resonance in this mass region.

A different interpretation of the enhancement of the cross section around 320 GeV was recently suggested in \cite{Maiani}, where the excess was linked to the second electroweak Higgs doublet required by low energy supersymmetry, and the CMS bump was tentatively identified with the second neutral Higgs boson of the MSSM.~\footnote{No  mention is made of the
   CDF events in \cite{Maiani}.} By contrast, the model of \cite{MN} avoids low energy supersymmetry altogether, but invokes conformal symmetry to explain and stabilize the electroweak hierarchy, postulating the absence of {\em any} intermediate scales between
the electroweak scale and the Planck scale. Confirmation of the properties outlined above (sharpness of the resonance, and branching ratios identical to those of the SM Higgs particle), together with the absence of any new fundamental fermions other than the right-chiral
neutrinos would constitute strong evidence for such a conformal scenario. At any rate,
it should  be relatively easy to discriminate between the present proposal and
alternative ones such as \cite{Maiani}, once enough statistics is accumulated.

\vspace{4mm}

\noindent
 {\bf Acknowledgment:} We are grateful to M. Kazana and P. Chankowski for discussions. We would also like to thank P.~Murat and A.~Robson for their comments on a first version of this letter.

\begin{figure}[ht]
\begin{minipage}[b]{0.49\linewidth}
\centering
\includegraphics[width=\textwidth,viewport=50 55 250 250,clip]{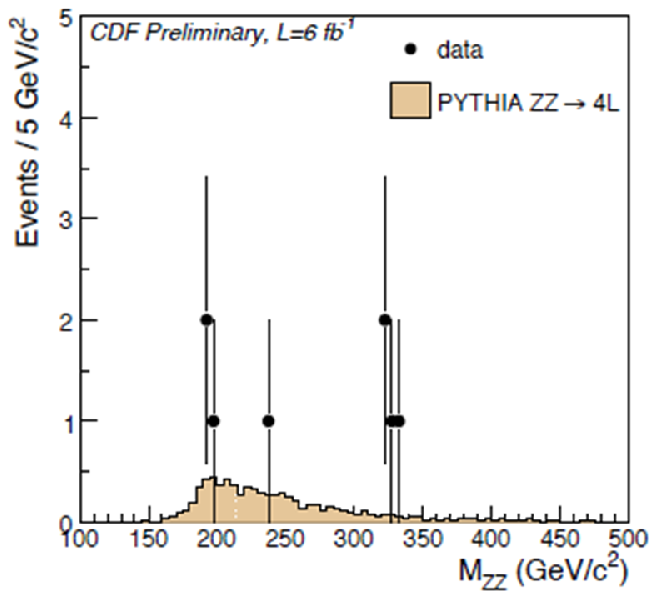}
\caption{The four lepton events reported by the CDF Collaboration \cite{CDF}.}
\label{fig:figure1}
\end{minipage}
\hspace{0.2cm}
\begin{minipage}[b]{0.45\linewidth}
\centering
\includegraphics[width=\textwidth,viewport=10 10 570 570,clip]{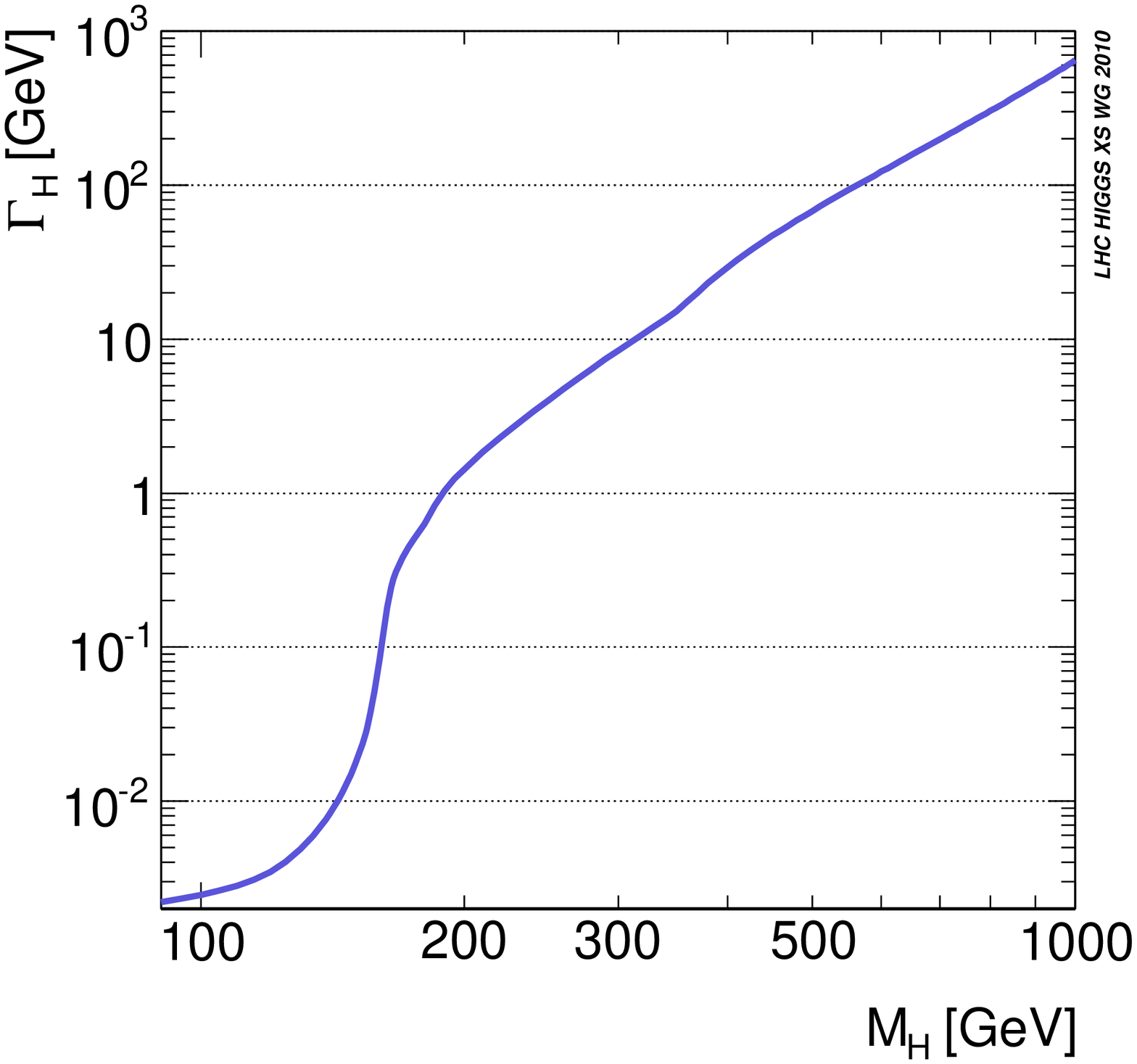}
\caption{$\GSM$ as a function of the SM Higgs mass \cite{Hdec}.}
\label{fig:figure2}
\end{minipage}
\end{figure}


\begin{thebibliography}{99}

\bibitem{CDF} CDF Collaboration, Phys.~Rev. {\bf D85} (2012) 012008, {\tt arXiv:1111.3432}

\bibitem{CMS} CMS Collaboration,
     Phys. Rev. Lett. 108, 111804 (2012),  {\tt arXiv:1202.1997[hep-ex]}

\bibitem{MN} K.A.~Meissner and H.~Nicolai,
{\it Conformal symmetry and the standard model},
Phys. Lett. {\bf B648} (2007) 312; {\tt hep-th/0612165},

\bibitem{CMSATLASHiggs} CMS Collaboration, Phys. Lett {\bf B716} (2012) 30; ATLAS Collaboration, Phys. Lett {\bf B716} (2012) 1.

\bibitem{BPY} W.~Buchm\"uller, R.D.~Peccei and T.~Yanagida, {\it  	
Leptogenesis as the origin of matter},
   Nucl. Part. Sci. {\bf 55} (2005) 311, {\tt arXiv:hep-ph/0502169}



\bibitem{MN2} K.A.~Meissner and H.~Nicolai, {\it Neutrinos, axions and conformal symmetry}, Eur.Phys. J. {\bf C 57} (2008) 493, {\tt arXiv:0803.2814[hep-th]}; A.~Latosinski, K.A.~Meissner and H.~Nicolai, {\it Neutrino Mixing and the Axion-Gluon vertex}, {\tt arXiv:1203.3886[hep-th]}

\bibitem{Hdec} LHC Higgs Cross Section Working Group, {\tt arXiv:1101.0593[hep-ph]}; A.~Djouadi, J.~Kalinowski, M.~Spira, {\it HDECAY}, Comput.Phys.Commun. 108 (1998) 56-74, {\tt arXiv:hep-ph/9704448}

\bibitem{Maiani} L.~Maiani, A.D.~Polosa and V.~Riquer, {\it Probing Minimal Supersymmetry at the LHC with the Higgs Boson Masses}, New J.Phys. {\bf 14} (2012) 073029, {\tt arXiv:1202.5998[hep-ph]}

\end{thebibliography}
\end{document}